%% file: ms.tex
\documentclass[preprint,12pt]{elsarticle}

\usepackage{amssymb}
\usepackage{xcolor}
\usepackage{blindtext}
\usepackage{import}
\usepackage{caption}
\usepackage{subfigure}
\usepackage{graphicx}
\usepackage{latexsym}
\usepackage{times}
\usepackage[version=3]{mhchem}
\usepackage{comment}
\usepackage{xspace,xcolor}
\usepackage{pdfpages}
\usepackage{float}
 \usepackage[labelfont=bf]{caption}
 \usepackage{amsmath}
 \usepackage[autostyle]{csquotes}
 \usepackage{xspace,xcolor}
\usepackage[capitalize]{cleveref}
\usepackage[a4paper, total={7in, 8in}]{geometry}
\usepackage{fancyhdr}
  \newcommand{\acknowledgement}
[1]
{
	\bgroup
	\flushleft
	\small\bf
	#1
	\par
	\egroup
}

\biboptions{sort&compress}

\journal{...}

\begin{document}


\begin{frontmatter}


\title{Numerical determination of iron dust laminar flame \\speeds with the counterflow twin-flame technique}

\author[First]{C.E.A.G van Gool\corref{cor1}}
\author[First]{T. Hazenberg}
\author[First]{J.A. van Oijen}
\author[First]{L.P.H. de Goey}

\cortext[cor1]{Corresponding author.\\c.e.a.g.v.gool@tue.nl (C.E.A.G. van Gool)}
\address[First]{Department of Mechanical Engineering, Eindhoven University of Technology, P.O. Box 513, NL-5600 MB, Eindhoven, the Netherlands}

\begin{abstract}
Iron dust counter-flow flames have been studied with the low-Mach-number combustion approximation. The model considers full coupling between the two phases, including particle/droplet drag. The dispersed phase flow strain relations are derived under the assumption of low Reynolds number conditions. The importance of solving a particle flow strain model is demonstrated by comparing three different models: a free unstrained flame, a counter-flow flame where particle flow strain is assumed equal to gas flow strain and one case in which the particle flow strain is solved. All three cases showed preferential diffusion effects, due to the lack of diffusion of iron in the fuel mixture, e.g. $D_{\ce{Fe},m}$ = 0. The preferential diffusion effect causes a peak in the fuel equivalence ratio in the preheat zone. At the burned side, the combined effect of strain and preferential diffusion showed a decrease in fuel equivalence ratio. Inertia effects, which are only included in the resolved particle flow strain case, counteract this effect and result in an increase of the fuel equivalence ratio at the burned side. A laminar flame speed analysis is performed and a recommendation is given on how to experimentally determine the flame speed in a counter-flow set-up. 
\end{abstract}


\end{frontmatter}

\section*{Novelty \& Significance}
We introduce a novel model to include particle flow strain in a dispersed counter-flow set-up. For the first time, the impact of particle strain on the flame structure of iron dust is studied.  Two major effects that modify the flame structure and burning velocity are identified: preferential diffusion and an inertia effect. Preferential diffusion effects are found to be always present in (iron) dust flames. The inertia effect plays a role in the resolved particle flow strain case. Due to the inertia of the particles, the particle flow strain is lower than the gas flow strain. As a consequence, higher particle concentrations are reached compared to the other cases. Furthermore, it is shown that each particle size experiences a different particle flow strain rate, which is important when doing experiments as it implies that the PSD at the flame front will be different than at the inlet.

\section*{Author Contribution}
C.E.A.G. van Gool: Conceptualization, Methodology, Software, Validation, Formal analysis, Investigation, Writing - Original Draft; T. Hazenberg: Conceptualization, Writing - Review \& Editing; J.A. van Oijen, L.P.H. de Goey: Conceptualization, Writing - Review \& Editing, Supervision, Funding acquisition. 

\input{Introduction}
\input{Methodology}
\input{Results}
\input{Conclusions}

\acknowledgement{Acknowledgments} \addvspace{10pt}
This project has received funding from the European Research Council (ERC) under the European Union' Horizon 2020 research and innovation program under Grant Agreement no. 884916.
\clearpage

\bibliography{biblio.bib}

\end{document}

%% file: Introduction.tex
\section{Introduction} \addvspace{10pt}
Recently a lot of research is performed to investigate the possibility of using iron powder as a zero-carbon fuel \cite{MIGNARD20075039, BERGTHORSON2015368, SUN19982405, TANG20091905, Hazenberg20214383, Ravi2022, VANGOOL2023}.  In closed-loop metal-fuel cycles, iron powder can be oxidized when power and/or heat is required. The formed oxides can be reduced at other places and other times using hydrogen, obtained via renewable energy sources. For the development of practical iron fuel-burning set-ups a thorough understanding on iron combustion characteristics is required. A key parameter for fundamental and practical purposes is the laminar burning velocity, $s_\mathrm{L}$.  

For gaseous flames it is well known that laminar flame speeds can be substantially modified by stretch effects such as flow non-uniformity, flame curvature, and flame/flow unsteadiness \cite{KARLOVITZ1953613,Lewis_Elbe1961,Markstein1964,WU1985,DIXONLEWIS1991305,deGoey1997}. These effects must be taken into consideration for an accurate prediction of the laminar flame speed. For that reason, Wu and Law \cite{WU1985} proposed the development of the counterflow twin-flame technique for the determination of $s_\mathrm{L}$. The first analytical studies of stretched flames were based on models with reduced complexity, with e.g. a single Lewis number, a single reversible reaction, and a flame sheet. Later on, the flame stretch theory was extended to more general flames, with finite flame front thickness \cite{deGoey1997,VANOIJEN201630}. For iron flames, stretch effects are also at play and additional complexity is expected when the fuel equivalence ratio is such that the condensed phase may comprise a substantial fraction of the total momentum of the flow, but not having the same velocity as the surrounding gas. 

When iron particles are exposed to a strained gas flow, like in the case of a counterflow, the gas is continuously accelerating tangential to the flame surface and so is the dispersed phase. The dispersed phase is dragged by the gas and therefore moves slower. A similar phenomenon is observed for coal particles flow and liquid droplet flow strained by gas flow \cite{CONTINILLO1990325,GRAVES19821189}. 

The goal of this paper is to quantify the effects of drag/stretch and preferential diffusion of general flame aerosol, with a focus on iron/air aerosols. A new model is derived to describe the particle flow strain as function of the gas flow strain under low particles Reynolds number conditions. The quasi-1D counterflow dispersed twin-flame configuration is based on the model of Hazenberg and van Oijen \cite{Hazenberg20214383}. With this model, the impact of stretch on the burning velocity is investigated. Furthermore, extrapolation to zero stretch (see e.g., \cite{WU1985}) is performed to obtain the unstretched laminair burning velocities. The particle strain model, does not limit itself to iron particles, it is also valid for other dispersed fuels that burn heterogeneously.

The derivation of a detailed and reduced one-dimensional (1D) particle flow strain model is presented in Section 2. Then, in the first part of Section 3 the preferential diffusion effects in a dispersed flame are discussed. In the second part, the influence of particle strain on flame structures is investigated. Next, particle flow strain effects for different particle sizes are investigated. In the last subsection of Section 3, two methods to measure the burning velocity are discussed. Finally, a discussion and the main conclusions are given in Section 4.


%% file: Methodology.tex
\section{Model description} 
\begin{figure}
	\centering
	\includegraphics[width=190pt]{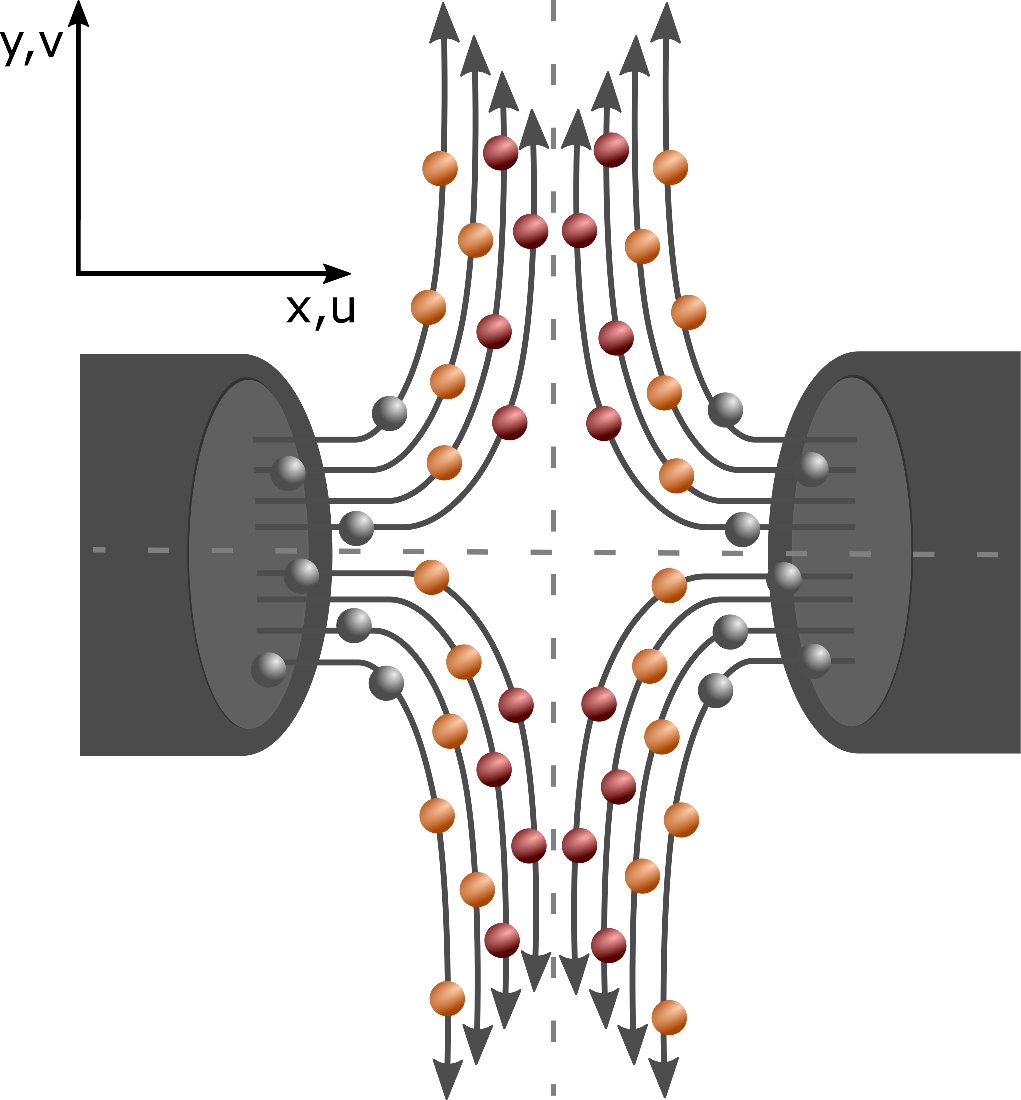}
	\caption{Schematic representation of a 1D counter-flow flame.}
	\label{flame_configuration}
\end{figure}

\addvspace{10pt}
The case of planar opposed flow nozzles, at a distance $L$ from each other, is considered. Compositions and velocities, at temperature $T_u$ = 300 K, are assumed uniform and equal at the nozzle exits. Therefore, the system is symmetric with respect to the stagnation plane. Under specific assumptions, which will be discussed in the remainder of this chapter, the system can be considered as quasi-1D. A schematic representation of the configuration is provided in Figure \ref{flame_configuration}. Due to symmetry at the stagnation plane, only half of the domain needs to be simulated. The iron-air flame is modeled with an Euler-Lagrange approach, similarly as in \cite{Hazenberg20214383,VANGOOL2023}.

\subsection{Particle model} 
The dispersed phase is modeled in a Lagrangian framework, where single particles are tracked in one-dimensional space such that the Lagrangian time-coordinate can be related to the Eulerian spatial-coordinate. The coupling between gas and dispersed phase is handled as in Sacomano Filho et al \cite{SacomanoFilho2018998}. The mass and enthalpy exchange between the phases are denoted as $S_m$ and $S_h$ and read
\begin{align}
	S_{m,k} = \dot{N}_{\mathrm{p},k}\frac{\Delta m_{\mathrm{p},k}}{V_k}, \quad S_{H,k} = \dot{N}_{\mathrm{p},k}\frac{\Delta H_{\mathrm{p},k}}{V_k}, 
\end{align}
with $\dot{N}_{\mathrm{p},k} = \dot{m}/m_{\mathrm{p,0}}$ the particle number flux, $\dot{m}$ the mass flux of particles at the inlet, $V_k$ the volume of computational cell $k$, and $\Delta m_{\mathrm{p},k}$ and $\Delta H_{\mathrm{p},k}$ the integrated mass and enthalpy changes of a single particle in a finite volume cell $k$. The equations for mass $m_\mathrm{p}$, enthalpy $H_\mathrm{p}$ and velocity in the x-direction $u_\mathrm{p}$ for a single particle are given by:
\begin{align}
	\frac{\mathrm{d}m_{\mathrm{p}}}{\mathrm{d}t} &= Y_{\ce{O2}}A_{\mathrm{d}}k_{\mathrm{d}}\mathrm{Da}^*,\\
	\frac{\mathrm{d}H_\mathrm{p}}{\mathrm{d}t} &= k_cA_{\mathrm{p}}\left(T_{\mathrm{g}}-T_{\mathrm{p}}\right) + \frac{\mathrm{d}m_{\mathrm{p}}}{\mathrm{d}t}h_{\ce{O2}},\\
	\frac{\mathrm{d}u_\mathrm{p}}{\mathrm{d}t} &= \frac{3}{4}\frac{C_{\mathrm{D}}\rho_{\mathrm{g}}}{d_{\mathrm{p}}\rho_{\mathrm{p}}}\Big|u_{\mathrm{g}}-u_{\mathrm{p}}\Big|\left(u_{\mathrm{g}}-u_{\mathrm{p}}\right), \\
	\frac{\mathrm{d}x_{\mathrm{p}}}{\mathrm{d}t} &= u_{\mathrm{p}}, 
\end{align}
where $Y_{\ce{O2}}$ is the mole fraction of oxygen in the carrier gas, $k_{\mathrm{d}}$ the mass-transfer coefficient for a sphere and Da$^*$ the normalized Damk\"oler number \cite{Hazenberg20214383}. The diffusive surface area is equal to the particle surface area, $A_{\mathrm{d}} = A_{\mathrm{p}}$, with $A_{\mathrm{p}} = \pi d^2_{\mathrm{p}}$. The heat transfer coefficient is given by $k_c$, the gas and particle temperature are $T_{\mathrm{g}}$ and $T_{\mathrm{p}}$ and the mass-specific enthalpy of the consumed oxygen $h_{\ce{O2}}$. Finally, $\rho_{\mathrm{p}}$ and $\rho_{\mathrm{g}}$ are the particle and gas density, $d_{\mathrm{p}}$ is the particle diameter, $x_{\mathrm{p}}$ the particle position and $C_\mathrm{D}$ the drag coefficient. The drag coefficient is formulated by an empirical relation for low Reynolds number \cite{clift2005bubbles}, 
\begin{align}\label{eq:drag}
	C_\mathrm{D} = \frac{24}{\mathrm{Re}}\left(1+0.15\mathrm{Re}^{0.687}\right).
\end{align} 
The particle Reynolds number $\mathrm{Re}$ is defined as, 
\begin{align}\label{eq:Re}
	\mathrm{Re} = \frac{d_{\mathrm{p}}|u_{\mathrm{g}}-u_{\mathrm{p}}|\rho_{\mathrm{g}}}{\mu_{\mathrm{g}}}
\end{align}
with $\mu_{\mathrm{g}}$ the dynamic viscosity of the carrier gas in the particle film layer. For further details of the particle model, we refer to the work of Hazenberg et al. \cite{Hazenberg20214383,Hazenberg2019}. As a first step, the oxidation beyond \ce{FeO} as presented in Gool et al \cite{VANGOOL2023} is not included here. 
\subsection{Particle strain model} \addvspace{10pt}
Consider a steady planar flame of finite thickness, such that the flame properties are only a function of the $x$-direction \cite{DIXONLEWIS1991305,CONTINILLO1990325}, i.e., $\rho(x)$, $T(x)$, $Y_i(x)$, etc. The only property which is a function of both the $x$- and $y$-direction is the velocity $v(x,y)$ in the $y$-direction. We assume that variations in the $y$-direction are small and at $y$~=~0, there is no velocity in the vertical direction, meaning $v(x,0)$~=~0.  Performing a first order Taylor expansion around the x-axis on the velocity in the $y$-direction results in
\begin{align}
	v(x,y) = y \frac{\partial v}{\partial y}\Big|_{y=0} = ay,
\end{align}
where we notice that $a(x) = \frac{\partial v}{\partial y}$ denotes the flow strain rate in the direction tangential to the flame surface, applicable for both particles and gas. Since the Taylor expansion is done around $y$ = 0, the strain can also be written as $a = v/y$, which will be useful later in the derivation. We now have a general expression for the strain, which is valid for both the gas and the particle flow strain. For the particle flow strain, we have to find expressions for $v_p$. The time derivative of the particle velocity in the $y$-direction can be described as,
\begin{align}\label{eq:v-particle-1}
	\frac{\mathrm{d}v_{\mathrm{p}}}{\mathrm{d}t} = \frac{3}{4}\frac{C_{D}\rho_{\mathrm{g}}}{d_{\mathrm{p}}\rho_{\mathrm{p}}}\big|v_{\mathrm{g}}-v_{\mathrm{p}}\big|\left(v_{\mathrm{g}}-v_{\mathrm{p}}\right)
\end{align}
with $v_{\mathrm{g}}$ the gas velocity in the $y$-direction. Substituting Eq.~\eqref{eq:drag} and Eq.~\eqref{eq:Re} in Eq.~\eqref{eq:v-particle-1} gives,
\begin{align}\label{eq:v-particle-2}
	\frac{\mathrm{d}v_{\mathrm{p}}}{\mathrm{d}t} = \frac{18\mu}{d^2_{\mathrm{p}}\rho_{\mathrm{p}}}\left(v_{\mathrm{g}}-v_{\mathrm{p}}\right)\left(1+0.15\mathrm{Re}^{0.687}\right).
\end{align}
We assume that $\mathrm{Re} \ll$ 1, which also follows from the Taylor expansion around $y$~=~0, such that we can linearize Eq.~\eqref{eq:v-particle-2} around $\mathrm{Re}$~=~0 which results in, 
\begin{align}\label{eq:v-particle-3}
	\frac{\mathrm{d}v_{\mathrm{p}}}{\mathrm{d}t} = \frac{18\mu_{\mathrm{g}}}{d^2_{\mathrm{p}}\rho_{\mathrm{p}}}\left(v_{\mathrm{g}}-v_{\mathrm{p}}\right) = \frac{v_{\mathrm{g}}-v_{\mathrm{p}}}{\tau}, && \text{with } \tau = \frac{d^2_{\mathrm{p}}\rho_{\mathrm{p}}}{18\mu_{\mathrm{g}}}.
\end{align}
Substituting $v_\mathrm{p} = 	\frac{\mathrm{d}y_{\mathrm{p}}}{\mathrm{d}t}$, a second order differential equation for particle position in $y$-direction $y_{\mathrm{p}}$ is obtained
\begin{align}\label{eq:v-particle-2nd-order}
	\frac{\mathrm{d}^2y_{\mathrm{p}}}{\mathrm{d}t} + \frac{1}{\tau}\frac{\mathrm{d}y_{\mathrm{p}}}{\mathrm{d}t} - \frac{a_{\mathrm{g}}}{\tau}y_{\mathrm{p}} = 0,
\end{align}
with $a_\mathrm{g} = \frac{v_{\mathrm{g}}}{y_{\mathrm{p}}}$ the gas strain at $y_{\mathrm{p}}$, the particle position. Notice that $y_{\mathrm{p}}$ is the only variable in Eq.~\eqref{eq:v-particle-2nd-order}, when assuming that $\tau$ and $a_g$ are constant.  Further assuming they are positive, the general solution of Eq.~\eqref{eq:v-particle-2nd-order} becomes
\begin{align}
	y_{\mathrm{p}} &= c_1 \exp\left(r_1t\right)+c_2\exp\left(r_2t\right), \qquad \text{with} \\
	r_1 &= -\frac{1}{2\tau}+\sqrt{\frac{1}{4\tau^2}+\frac{a_{\mathrm{g}}}{\tau}},  &&r_2 = -\frac{1}{2\tau}-\sqrt{\frac{1}{4\tau^2}+\frac{a_{\mathrm{g}}}{\tau}}.\nonumber \\
	r_1 &> 0,  &&r_2 <0
\end{align}

Then, by using the definition of strain and substituting the general solution for particle position and its derivative, an expression can be obtained for the particle strain for $t \rightarrow \infty$ when the strain has reached equilibrium:
\begin{align}
	a_\mathrm{p, eq} &= \frac{v_{\mathrm{p, eq}}}{y_{\mathrm{p, eq}}} \\
	 &= \frac{c_1r_1 \exp\left(r_1t\right)+c_2r_2\exp\left(r_2t\right)}{c_1 \exp\left(r_1t\right)+c_2\exp\left(r_2t\right)}\Big|_{t\rightarrow\infty} = r_1. 
\end{align}
For $t \rightarrow \infty$, the terms with $r_2$ go to zero. $r_2$ can physically be interpreted as the adjusting rate of the particle flow field when subjected to a continuous gas flow strain. In a continuously strained flow field, the particle is not able to follow the accelerating gas due to inertia. Now an expression can be derived for the change in particle flow strain, using $a_\mathrm{p, eq}, a_\mathrm{p}$ and $r_2$, assuming that strain reaches the equilibrium state within time constant $\tau_{r_2} = -1/r_2$, 
\begin{align}\label{eq:a-particle-reduced}
	\frac{\mathrm{d}a_{\mathrm{p}}}{\mathrm{d}t}= \frac{a_{\mathrm{p, eq}} - a_{\mathrm{p}}}{\tau_{r_2}}.
\end{align}
The particle flow strain acts as a sink term ($a_{\mathrm{p}} \geq 0$ and $a_{\mathrm{g}} > 0$ in counterflow flames) on the particle numberflux $\dot{N}$. This means that an additional equation for the change in particle numberflux $\dot{N}_{\mathrm{p}}$
\begin{align}\label{eq:nflux}
	\frac{\mathrm{d} \dot{N}_{\mathrm{p}}}{\mathrm{d} t} = -a_{\mathrm{p}}\dot{N}_{\mathrm{p}}
\end{align}
is solved.

\subsection{Gas-phase modeling}
The counterflow-twin flame is modeled as a steady quasi-1D stretched flame, where the low-Mach number combustion-approximation is assumed \cite{law_2006}. The finite volume solver CHEM1D is used to solve these equations \cite{Somers1994}. The quasi-1D conservation equations, read:
\begin{align}
	\frac{\partial}{\partial x} \left( \rho u \right) &= S_m - \rho a_{\mathrm{g}} \\ \label{eq:species_conservation}
	\frac{\partial}{\partial x} \left( \rho u Y_i\right) +\frac{\partial}{\partial x}\left( \rho U_{i} Y_i\right) &= \delta_{i,k}S_{m}- \rho a_{\mathrm{g}}Y_i \\ 
	 \frac{\partial }{\partial x}\left( \rho uh\right) + \frac{\partial q}{\partial x} &= S_h -  \rho a_{\mathrm{g}} h
\end{align}
with $U_{i}$ the diffusion velocity of species $i$. Since oxygen and nitrogen are the only species in the gas phase, and nitrogen is used to ensure conservation of mass, the $i$ in Eq.~\eqref{eq:species_conservation} equals the index of oxygen. As there is only exchange of oxygen, the $k$ in the Kronecker delta $ \delta_{i,k}$ equals the index of oxygen. The transport of enthalpy due to mass diffusion and conduction in the gas phase as
\begin{align}
	q =  - \frac{\lambda}{c_p} \frac{\partial h}{\partial x} +  \sum_{i=1}^{N_s}h_i \left(\frac{\lambda}{c_{\mathrm{p}}}\frac{\partial Y_i}{\partial x} - \rho U_i Y_i\right).
\end{align}

In this 1-D formulation, $a_{\mathrm{g}}$ denotes the velocity stretch rate field in the direction tangential to the flame surface $y$, which is given by an expression similar to that for the particle flow strain 
\begin{align}
	a_{\mathrm{g}}(x) = \frac{\partial v}{\partial y}.
\end{align} 
which is obtained through solving an additional transport equation:
\begin{align}
	\frac{\partial \left(\rho ua_\mathrm{g}\right)}{\partial x} - \frac{\partial}{\partial x}\left(\mu \frac{\partial a_\mathrm{g}}{\partial x}\right) = S_{a} + J - \rho a^2
\end{align}
where $S_a = \dot{N}_{\mathrm{p},k} \frac{\Delta m_{\mathrm{p},k}}{V_k}$ is the exchange of strain between dispersed and gas phase, $J = \frac{1}{y}\frac{\partial p}{\partial y}$ is the tangential pressure gradient.

\subsection{Boundary conditions}
There are two sets of boundary conditions possible to solve the flow strain in the setup shown in Fig. \ref{flame_configuration}. One can either prescribe the strain rate and solve for the corresponding mass-fluxes, or the mass-flux on the inlet is prescribed and the strain field is solved for. The latter is used in this work. Since only half of the domain is modeled, a mirror boundary condition is used at $x$ = 0 for all properties of the gas phase. The tangential pressure gradient $J$ is constant throughout the flow field, and an eigenvalue of the system. As it is expected that only a small fraction of particles cross the stagnation plane, no mirror boundary condition for the particles is used, instead they are removed when they cross x=0. 

%% file: Results.tex
\section{Results} \addvspace{10pt}
\begin{figure}
	\centering
	\includegraphics[width=400pt]{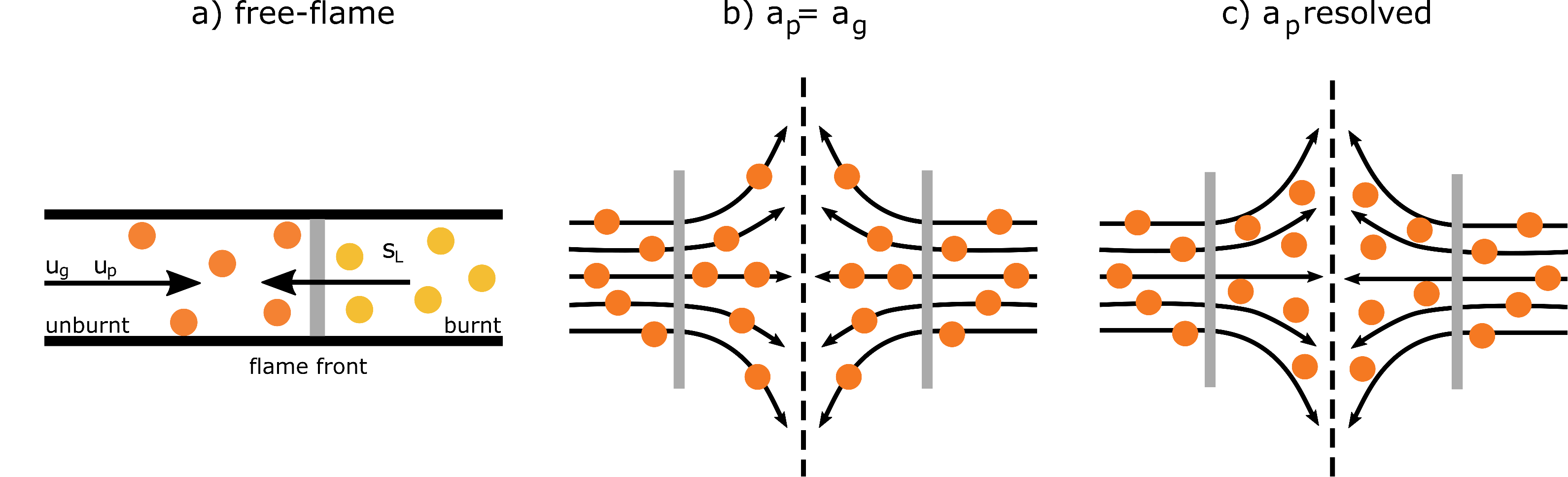}
	\caption{Schematic representation of a) a one-dimensional planar flame, b) a stagnation flame with $a_{\mathrm{p}}$~=~$a_{\mathrm{g}}$, and c) a stagnation flame with $a_{\mathrm{p}}$ solved according to Eqs.~(\ref{eq:a-particle-reduced},~\ref{eq:nflux}). Black arrows are direction of the gas velocity, grey lines represent the flame front and the dashed lines are the location of the stagnation plane.}
	\label{fig:cases1-3}
\end{figure}
In this section, the influence of particle strain, described by Eq.~\ref{eq:a-particle-reduced}, on the flame structures is investigated. The influence of particles and the combination of particles and strain on thermo-diffusivity and preferential diffusion are not yet well understood. For that purpose, three cases are employed (1) a free unstrained flame where $a_{\mathrm{p}}$~=~$a_{\mathrm{g}}$ = 0 and without slip ($u_{\mathrm{p}}$~=~$u_{\mathrm{g}}$), (2) an opposed twin flame where $a_{\mathrm{p}}$~=~$a_{\mathrm{g}}$ and (3) an opposed twin flame where the particle strain is solved according to Eqs.~(\ref{eq:a-particle-reduced},~\ref{eq:nflux}), a schematic representation of these cases is shown in Fig. \ref{fig:cases1-3}. 

\subsection{Preferential diffusion effects in a dispersed free flame}
\begin{figure}
	\centering
	\subfigure[]{\includegraphics[width=0.48\textwidth]{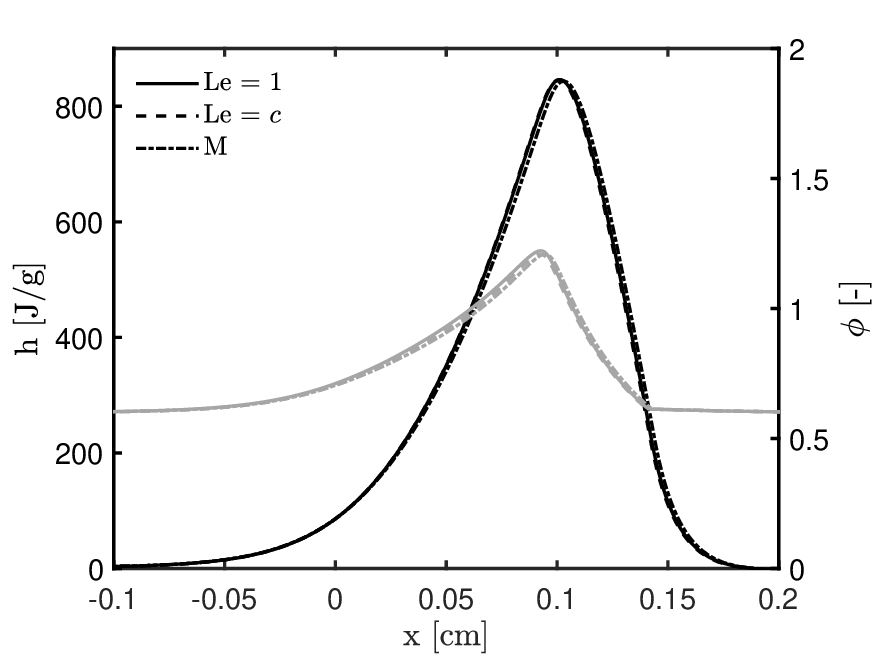}\label{fig:h-phi}}
	\quad
	\subfigure[]{\includegraphics[width=0.48\textwidth]{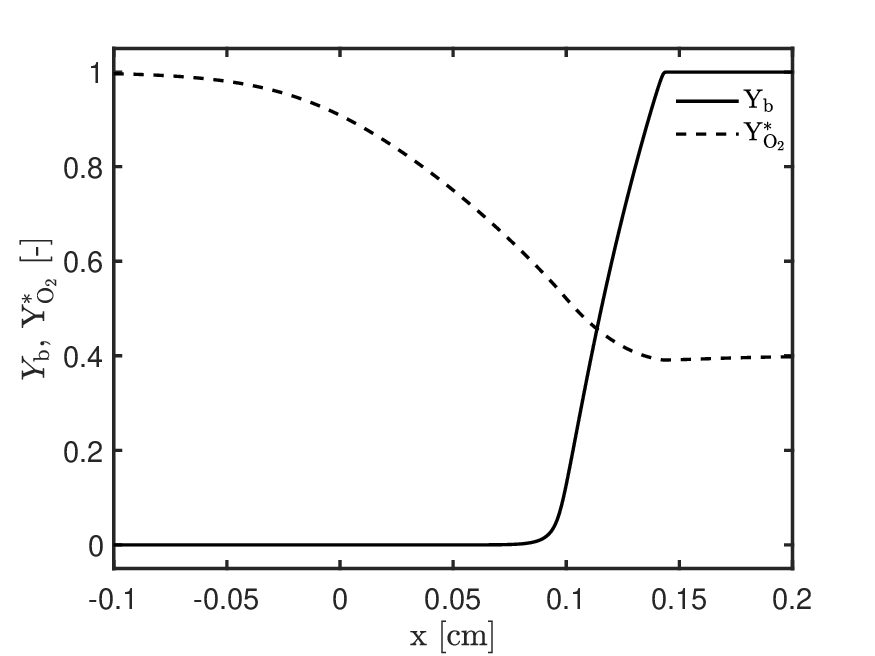}\label{fig:Yb-YO2}}
	\caption{a) Profile of enthalpy (black) and $\phi$ (grey) through a one-dimensional lean ($\phi$ = 0.6) iron-dust/air flame without flame stretch and unity Lewis numbers (solid), constant non-unity Lewis numbers (dashed) and mixture averaged diffusion coefficient (dash-dotted). b) Profile of $Y_{\mathrm{b}}$ (solid) and $Y^*_{\ce{O2}}$ (dashed) at $\phi$ = 0.6 and Le = 1.}		
	\label{fig:x-h_tot}
\end{figure}
The freely propagating flame configuration as in \cite{Hazenberg20214383,VANGOOL2023}, is a stretchless flame. When no flame stretch ($a_{\mathrm{p}}$~=~$a_{\mathrm{g}}$ = 0) as well as no preferential diffusion (Le$_i$~=~1) effects are at play in a gaseous flame, enthalpy and element mass fraction are constant through the flame \cite{Swart2009}. However, in Fig.~\ref{fig:h-phi} it can be seen that the enthalpy and fuel equivalence ratio $\phi = \frac{Z_{\ce{Fe}}}{Z_{\ce{O}}} \cdot s$, with $Z_{\ce{Fe}}$ and $Z_{\ce{O}}$ the element fraction of iron and oxygen in both solid and gas phase, and $s$ the stoichiometric ratio assuming \ce{FeO} as final product, are not constant through the flame for a dispersed flame where gaseous species have unity Lewis numbers. This can be explained by the fact that iron in the particles cannot diffuse, meaning $D_{\ce{Fe},m} $~=~0, resulting in Le$_{\ce{Fe}}$~=~$\infty$. Also in the work of Wright et al. \cite{WRIGHT2016} it is mentioned that the mass diffusivity of the particles is practically absent and therefore an infinite Lewis number for the dispersed phase is to be expected. This indicates that preferential diffusion effects will always be present, when dealing with non-volatile dispersed phase combustion. The enthalpy and $\phi$ for a constant, non-unity Lewis number and mixture averaged diffusion coefficient are also shown in Fig.~\ref{fig:h-phi}. Only small deviations are present compared to the (Le$_i$ = 1) case, because the Lewis number of oxygen is close to unity and the only gas phase species with Le$_i$. In the remainder of this work a mixture averaged diffusion coefficient for \ce{O2} is used. Fig. \ref{fig:Yb-YO2} shows the burned fraction $Y_{\mathrm{b}} = 1 - \frac{m_{\mathrm{p, u}}}{m_{\mathrm{p}}}$ of the particles and the normalized oxygen fraction $Y^*_{\ce{O2}}$ = $Y_{\ce{O2}}/Y_{\ce{O2}\mathrm{,0}}$. As already observed in the work of Hazenberg and van Oijen \cite{Hazenberg20214383}, oxygen diffuses to the reaction front such that the normalized oxygen fraction starts to decrease before the particle burned fraction starts to decrease. This explains the local increase in $\phi$ and may result in local fuel-rich conditions even when inlet conditions are lean.

\subsection{Flame structures}\label{Result:compare}
\begin{table}[h!] \small \caption{Simulation parameters.}
	\centerline{\begin{tabular}{llll}		
			\hline 
			Variable & Value & Unit     \\
			\hline
			$\rho_{\mathrm{Fe}}$     & 7.874	& $\mathrm{g/cm^3}$ \\
			$\rho_{\mathrm{FeO_x}}$  & 5.745	& $\mathrm{g/cm^3}$ \\ 
			$k_{\infty}$             & 75.0 $\times 10^7$ & cm/s         \\ 
			$T_\mathrm{a}$           & 14.4 $\times 10^3$ & K   		 \\
			$c_\mathrm{p,p}$	     & 0.76       & J/gK      \\
			$\Delta h_\mathrm{c}$ 	 & 4550  & J/g       \\
			\hline 
	\end{tabular}}
	\label{tab:paramters}
\end{table}

In this subsection, flame structures are compared of the three cases depicted in Fig. \ref{fig:cases1-3}. For the particles, the parameters from Table \ref{tab:paramters} are used. A mono-dispersed powder with initial size $d_{\mathrm{p,0}}$~=~10 \textmu m is utilized. The fuel equivalence ratio at the inlet is set to $\phi$~=~0.6. The mass-flux of gas at the inlet for cases (2) and (3) is $\dot{m}_{\mathrm{g}}$~=~0.04~g/cm$^2$s. For the free flame, case (1), there is no initial mass-flux prescribed as for this case the mass-flux at the inlet is an eigenvalue problem. The $x$-axis values in all graphs in this section are not the actual position of the counter-flow flames. Instead, the counter-flow flames are shifted in position such that the location of the maximum temperature is coincides with that of the free flame. 

\begin{figure}
	\centering	
	\subfigure[]{\includegraphics[width=0.48\textwidth]{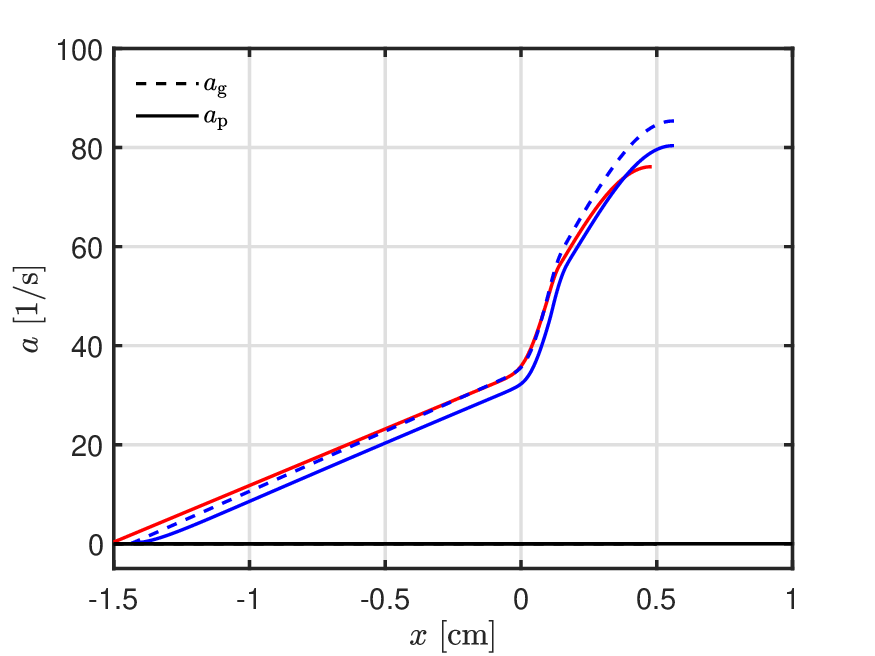}\label{fig:K}}
	\quad
	\subfigure[]{\includegraphics[width=0.48\textwidth]{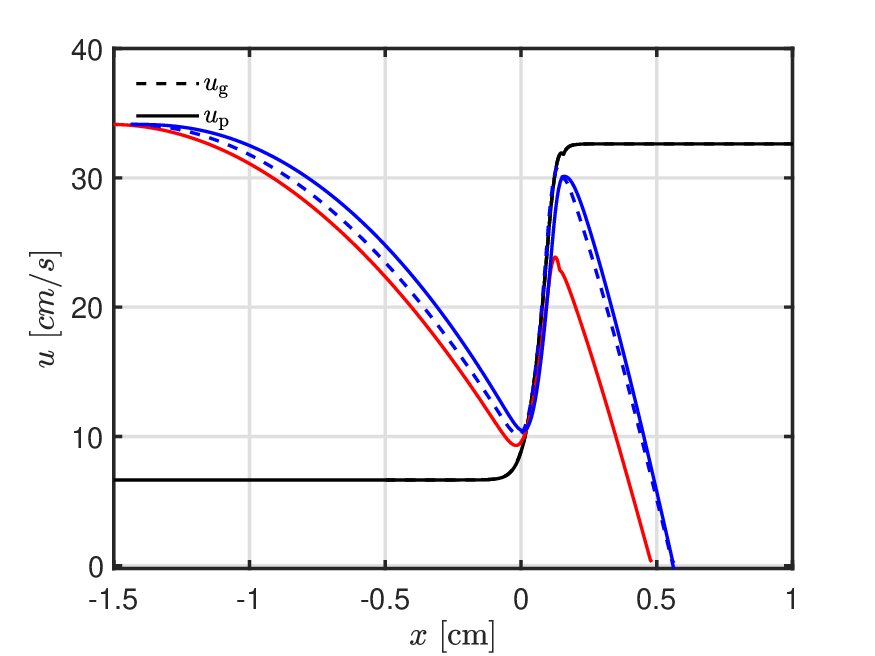}\label{fig:u}}		
	\caption{(1) Free flame (black), (2) $a_{\mathrm{p}}$ = $a_{\mathrm{g}}$ (red) and (3) resolved $a_{\mathrm{p}}$ (blue). a) Profile of particle (solid) and gas (dashed) flow stretch. b) Velocity profile of gas (dashed) and particle (solid). Case (2) and (3) at $\phi$ = 0.6 and $\dot{m}$ = 0.04 g/cm$^2$s} \label{fig:flame_structures_K-u}
\end{figure}
In Fig. \ref{fig:flame_structures_K-u}, the gas and particle flow strain and velocity profile of all three cases are shown. The flow strain of the free flame (black) is equal to zero as this is a stretch-free case. The strain profiles of the $a_{\mathrm{p}}$~=~$a_{\mathrm{g}}$ (red) and resolved $a_{\mathrm{p}}$ (blue) cases are deviating from each other, especially in terms of maximum strain rate. Both cases have an inlet $\dot{m}_{\mathrm{g}}$ = 0.04 g/cm$^2$s, but since the particle flow and gas flow in the  $a_{\mathrm{p}}$~=~$a_{\mathrm{g}}$ case basically move as one inertia, being different from the resolved $a_{\mathrm{p}}$ case where the gas flow accelerates the particle flow and the particle flow decelerate the gas flow, different strain rates are obtained throughout the domain. 

\begin{figure}
	\centering	
	\subfigure[]{\includegraphics[width=0.48\textwidth]{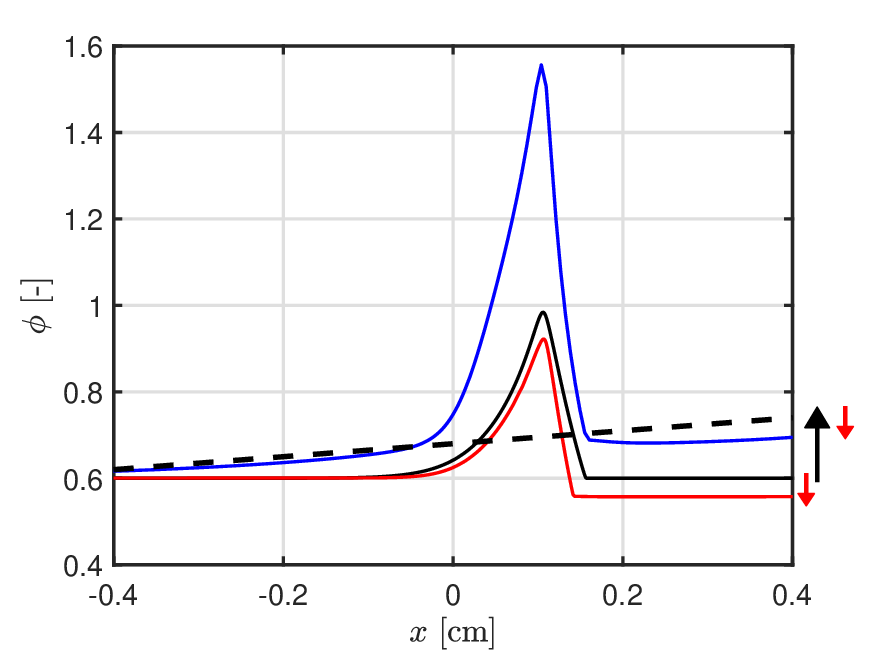}\label{fig:phi}}
	\quad
	\subfigure[]{\includegraphics[width=0.48\textwidth]{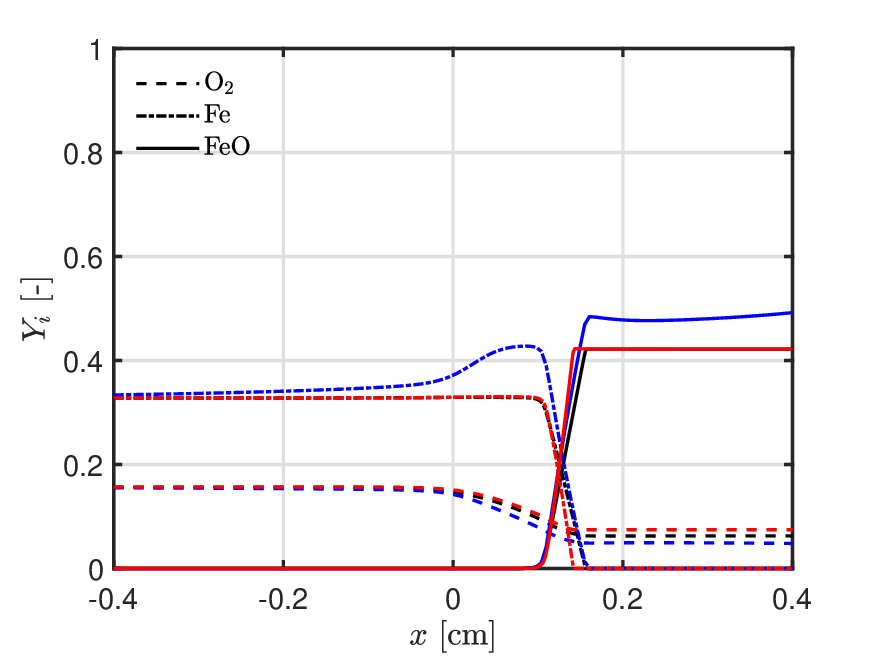}\label{fig:Yi}}	
	\quad
	\subfigure[]{\includegraphics[width=0.48\textwidth]{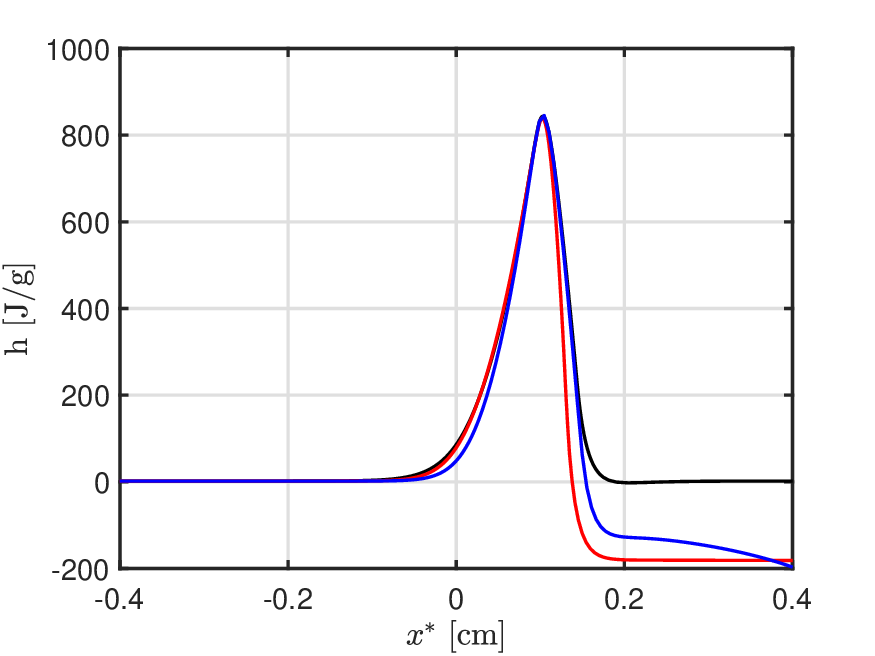}\label{fig:h}}	
	\quad
	\subfigure[]{\includegraphics[width=0.48\textwidth]{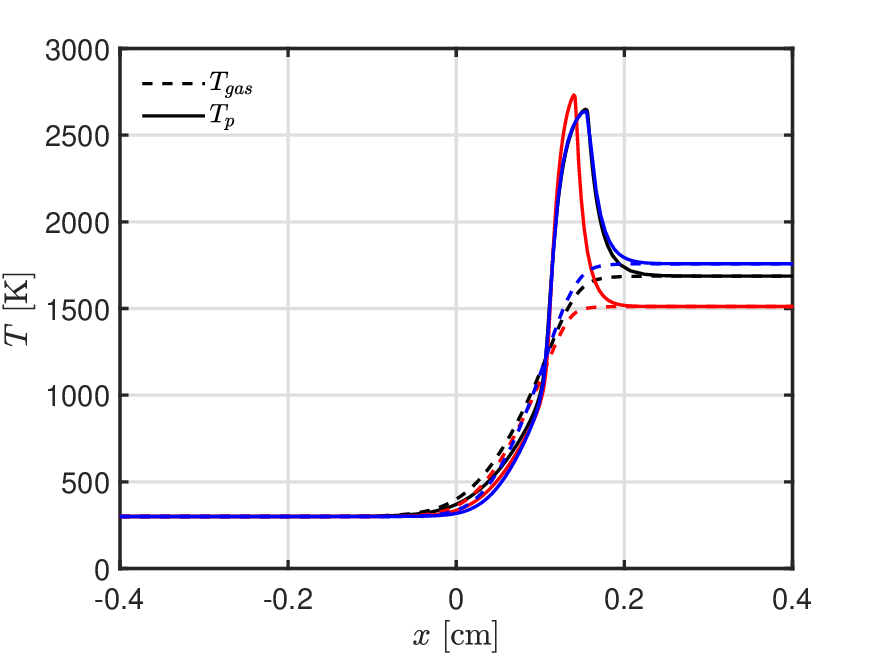}\label{fig:T}}
	\caption{(1) Free flame (black), (2) $a_{\mathrm{p}}$ = $a_{\mathrm{g}}$ (red) and (3) resolved $a_{\mathrm{p}}$ (blue). a) fuel equivalence ratio, where the dashed line is added for demonstrative purposes to show the influence of strain~+~inertia, which already plays a role on the unburned side. The black arrow indicates the influence of strain~+~inertia at the burned side and the red arrows represent the influence of strain~+~preferential diffusion at the burned side. b) species mass fractions. c) specific enthalpy profile. d) temperature profile of particle (solid) and gas (dashed). Case (2) and (3) at $\phi$ = 0.6 and $\dot{m}$ = 0.04 g/cm$^2$s} \label{fig:flame_structures_T-h-phi-Y}
\end{figure}

Figure \ref{fig:u} displays the particle and gas velocity, where it can be seen that slip effects are at play for case (3). For cases (1) and (2) slip effects are not present as $a_{\mathrm{p}}$~=~$a_{\mathrm{g}}$ and also $u_{\mathrm{p}}$ = $u_{\mathrm{g}}$. For case (3) it is observed that initially, as gas and particles move through the domain, the gas decelerates a bit faster than the particles. As soon as the particles start to burn and temperature rises, see Fig. \ref{fig:flame_structures_T-h-phi-Y},  the velocities will increase and the gas velocity will slightly exceed the velocity of the particles. After that, particle and gas velocity decrease again resulting in particles having a somewhat higher velocity than the gas. The effect of particles being pulled by the gas flow, not being able to keep up with the gas flow and therefore decelerating the gas flow is similar to the accelerating and decelerating effects as seen in Fig. \ref{fig:K}. 
	
In Fig. \ref{fig:flame_structures_T-h-phi-Y}, profiles of fuel equivalence ratio, mass fraction, temperature and specific enthalpy are shown. The equivalence ratio of the free flame is constant across the flame ($\phi_{\mathrm{u}}$~=~$\phi_{\mathrm{b}}$) as can be seen in Fig. \ref{fig:phi}. There are however local deviations due to preferential diffusion. In the preheat-zone, before any oxygen is consumed, oxygen gas has a higher velocity, because of diffusion to the reaction layer, which results in a decrease of oxygen concentration, see Fig. \ref{fig:Yi}. Due to the decrease in oxygen mass fraction, the fuel equivalence ratio increases. As soon as the particles start to consume oxygen, the fuel equivalence ratio decreases again. The fuel equivalence ratio profile of the $a_{\mathrm{p}}$ = $a_{\mathrm{g}}$ case also shows this preferential diffusion effect, indicated by the local increase in $\phi$. This effect combined with flow straining results in relatively less straining of oxygen and therefore a decrease in fuel equivalence ratio at the burned side. The red arrows indicate the effect of strain and preferential diffusion. The fuel equivalence ratio of the resolved $a_{\mathrm{p}}$ case increases already before the preheat-zone, due to the combined effect of strain and inertia of the particles. This effect results in an enrichment of the fuel equivalence ratio as demonstrated by the black dashed line and the black arrow. On top of that there is the effect of strain and preferential diffusion, which has a decreasing influence on the fuel equivalence ratio. However, the inertia effect is dominant and therefore the fuel equivalence ratio at the burned side increases. 

The specific enthalpy of the free flame is constant across the flame ($h_\mathrm{u}$~=~$h_{\mathrm{b}}$) as can be seen in Fig.~\ref{fig:h}. A peak in specific enthalpy occurs due to preferential diffusion. The $a_{\mathrm{p}}$ = $a_{\mathrm{g}}$ case shows a similar peak due to preferential diffusion. There is a significant amount of leaking where the specific enthalpy is higher, see Fig.~\ref{fig:K}, such that a relative large amount of enthalpy is transported. Therefore, a lower specific enthalpy is observed at the burned side compared to the free flame. For case (3) we also observe a peak in specific enthalpy due to preferential diffusion and a decrease at the burned side due to the combined effect of strain and preferential diffusion. The latter effect seems lower than for case (2), due to the combined effect of strain and inertia. We also see that instead of reaching a plateau, which we do observe for case (1) and case (2), the specific enthalpy keeps decreasing. This is because the fuel equivalence ratio is still increasing due to faster leakage of the gas phase, indicating that near the stagnation plane the inertia effect will dominate.

The temperature profiles, shown in Fig.~\ref{fig:T}, of the three cases are not that different from each other until the maximum temperature is reached. Even though the specific enthalpy at the burned side has decreased, this does not mean that the flame temperature reduces, as can be seen in Fig.~\ref{fig:T}. One should keep in mind that the heat capacity of solid iron oxide is significantly lower, see Table \ref{tab:paramters}, than the specific heat capacity of oxygen ($c_\mathrm{p,\ce{O2}} >$  1 J/gK). Since the iron particles contribute to about half of the total mass at the end of the domain, see Fig. \ref{fig:Yi}, the effective heat capacity of the mixture is severely lowered. Therefore, it is possible for the temperature to rise while the specific enthalpy decreases. 

\subsection{Particle flow strain for different particle sizes}\label{Result:dp}
\begin{figure}
	\centering	
	\subfigure[]{\includegraphics[width=0.48\textwidth]{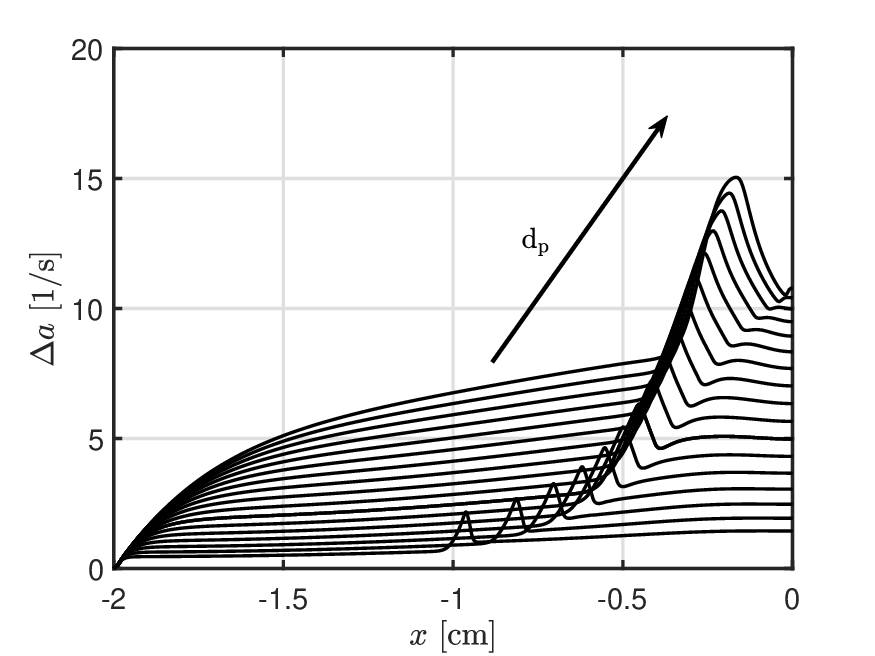}\label{fig:dp-K}}	
	\quad
	\subfigure[]{\includegraphics[width=0.48\textwidth]{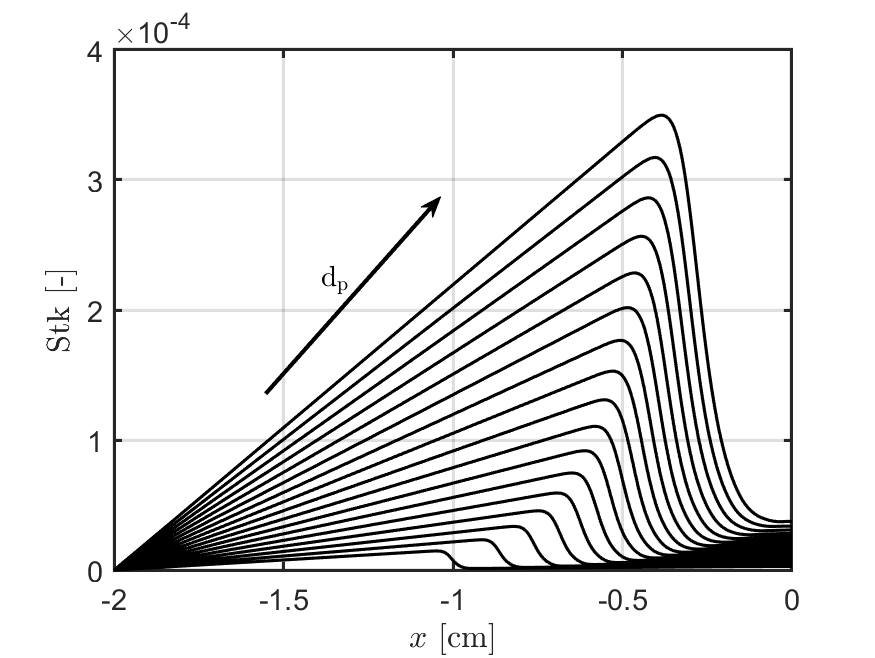}\label{fig:dp-Stk}}
	\caption{a) Absolute difference between gas flow strain and particle strain ($\Delta a$ = $a_{\mathrm{p}}$ - $a_{\mathrm{g}}$). b) Stokes number profile. Simulations are performed at $\phi$ = 0.6 and $\dot{m}$ = 0.04 g/cm$^2$s for particle sizes of $d_{\mathrm{p}}$~=~4:1:20 \textmu m .} \label{fig:dp_flame_structures_T-h-phi-Y}
\end{figure} 

\begin{figure}
	\centering
	\includegraphics[width=260pt]{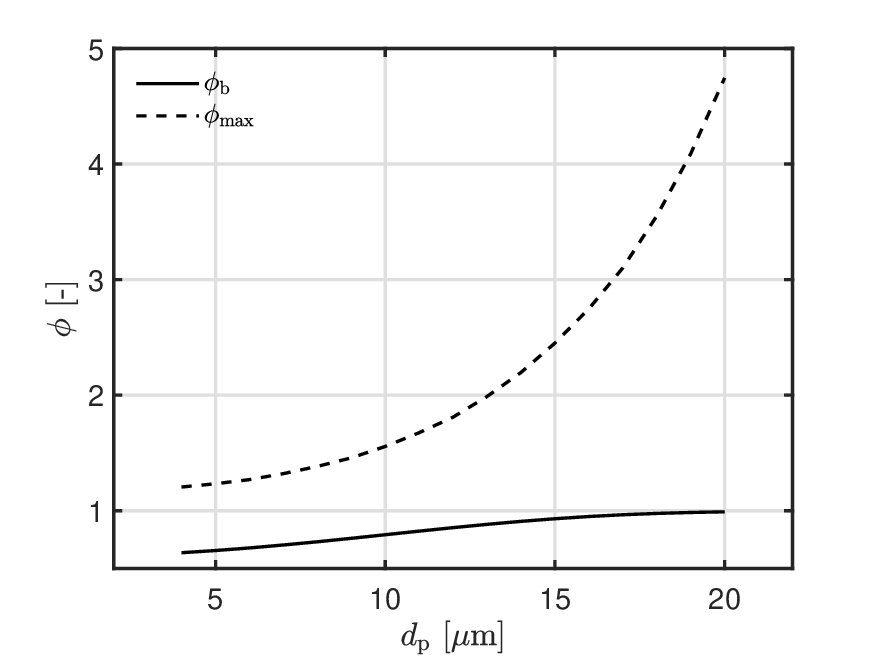}
	\caption{Burned and maximum values of fuel equivalence ratio as function of particle size. Simulations are performed at $\phi$ = 0.6 and $\dot{m}$ = 0.04 g/cm$^2$s for particle sizes of $d_{\mathrm{p}}$~=~4:1:20 \textmu m .}
	\label{fig:dp_flame_structures_phi}
\end{figure}
In the previous section we have shown the importance of solving a separate model for the particle flow strain. In this section we will use that model to explore the influence of strain when using various particle sizes. 

The absolute difference between gas and particle flow strain $\Delta a$ for particle sizes of $d_{\mathrm{p}}$ = 4-20 \textmu m with an interval of 1 \textmu m are shown in Fig. \ref{fig:dp-K}. As the particle size increases, the particles are less capable of following the gas flow, indicated by the increase in the Stokes number (Stk = $\tau_{\mathrm{p}} a_{\mathrm{g}}$) for larger particles in Fig.~\ref{fig:dp-Stk}. As a consequence, the combined effect of strain and inertia becomes more significant and the absolute difference in strain increases for larger particles. The increase in absolute strain difference results in an increase in both maximum and burned $\phi$, as can be seen in Fig.~\ref{fig:dp_flame_structures_phi}. This implies that, when performing counter-flow experiments with a particle distribution, each particle experiences a different \ce{O2} concentration at the flame front. 

\subsection{Laminar flame speed prediction}\label{Result:sL}
\begin{figure}
	\centering	
	\subfigure[]{\includegraphics[width=0.48\textwidth]{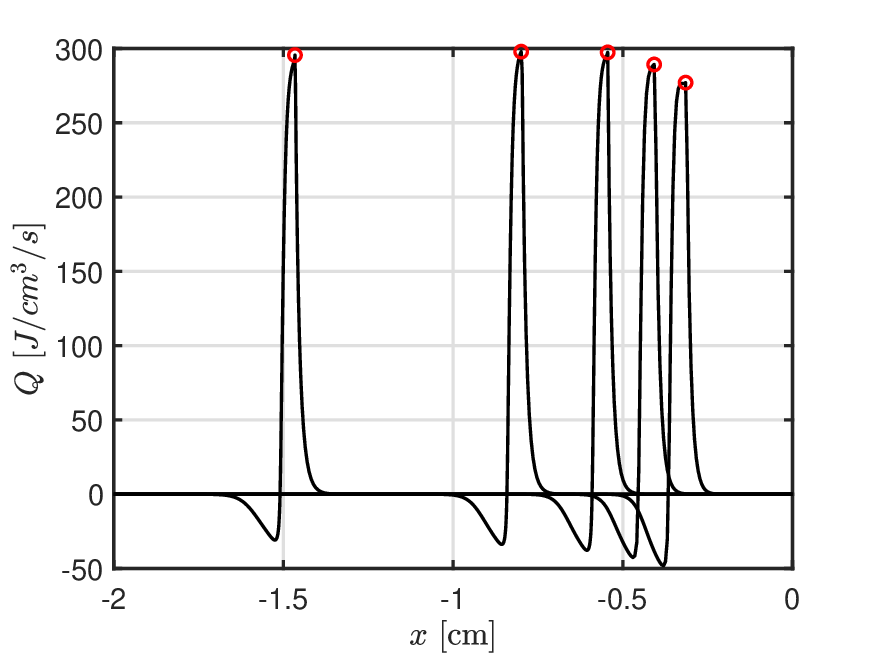}\label{fig:Q-Qmax}}
	\quad
	\subfigure[]{\includegraphics[width=0.48\textwidth]{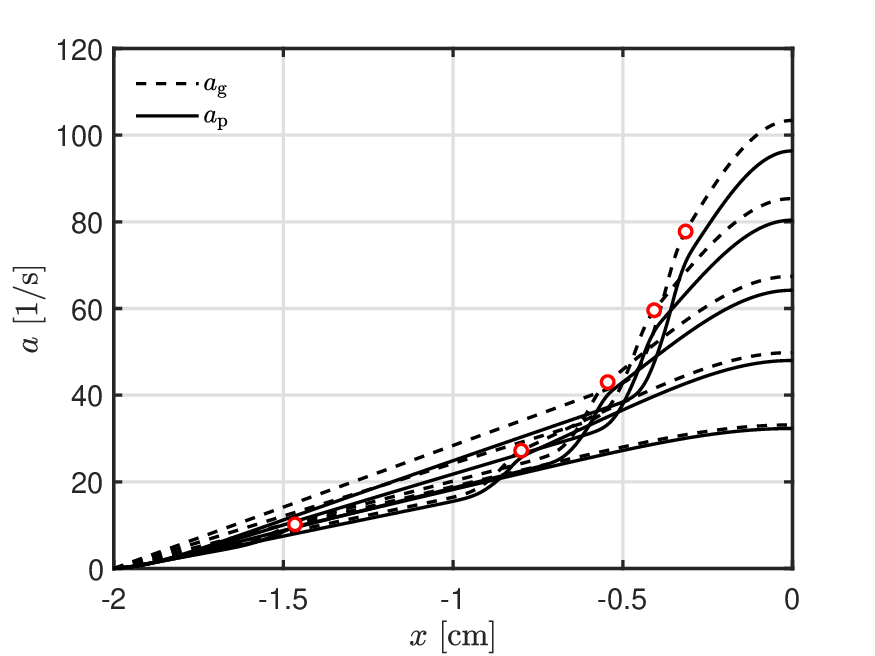}\label{fig:strain-Qmax}}	
	\quad
	\subfigure[]{\includegraphics[width=0.48\textwidth]{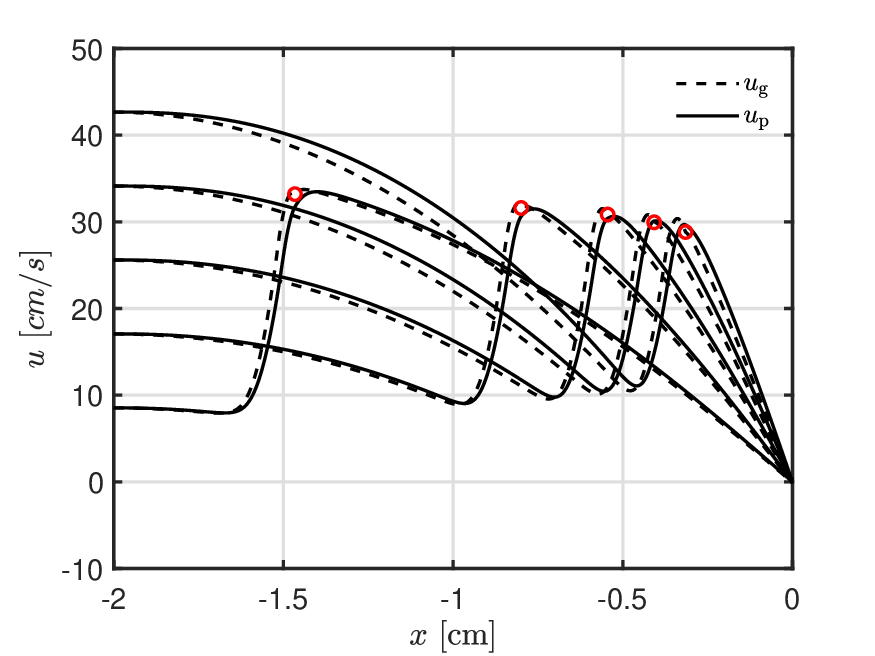}\label{fig:u_ref-Qmax}}
	\quad
	\subfigure[]{\includegraphics[width=0.48\textwidth]{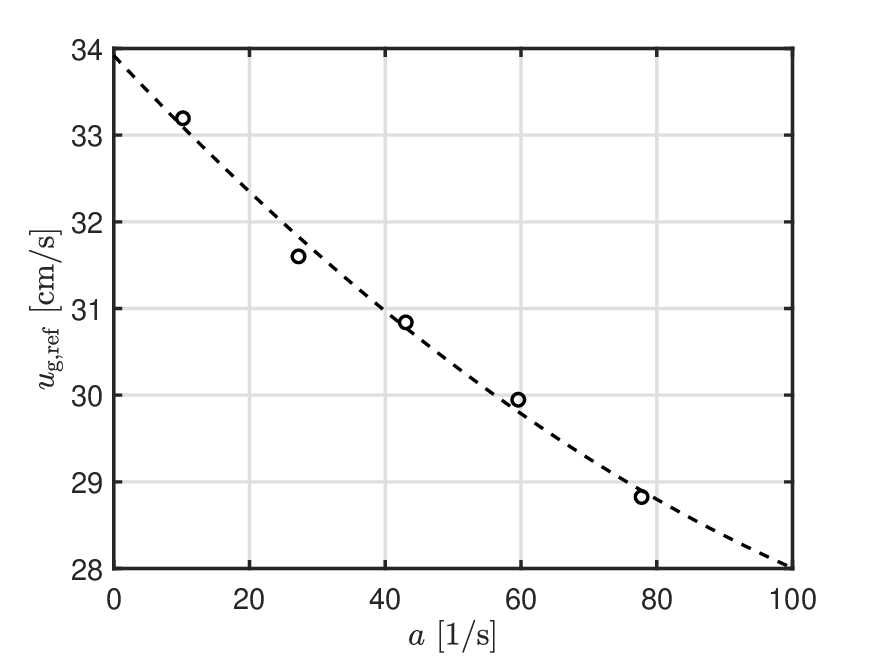}\label{fig:K-sL-Qmax}}	
	\caption{$s_{\mathrm{L}}$ determination at the location of $Q_{\mathrm{max}}$. a) Heat release rate profiles. b) Gas (dashed) and particle (solid) strain profiles. c) Gas (dashed) and particle (solid) velocity profiles. d) $u_{\mathrm{ref}}$ as function of strain rate. Simulations are performed at $\phi$~=~0.6 and $\dot{m}$~=~0.01~-~0.05 g/cm$^2$s, with an interval of 0.01~g/cm$^2$s. The red symbols indicate the location of $Q_{\mathrm{max}}$.} \label{fig:sL-Qmax}
\end{figure} 

There are several ways to perform flame speed predictions \cite{WU1985,LAW1989,TIEN1991}. Here we numerically employ two variants of a similar method \cite{WU1985}. One of the options is the maximum heat release variant, where reference gas velocity and strain rate are used at the location of maximum heat release, $Q_{\mathrm{max}}$, as the name of this variant already gives away. In Fig. \ref{fig:sL-Qmax} the heat release rate, strain rate, velocity profiles and the effect of strain on the gas velocity at the location of maximum heat release, $u_{\mathrm{g, ref}}$, for $\phi_{\textrm{in}}$~=~0.6 and mass-flux $\dot{m}_{\mathrm{g}}$~=~0.01~-~0.05 g/cm$^2$ are shown. The red dots in Fig.~\ref{fig:Q-Qmax}, indicate the position of $Q_{\mathrm{max}}$, which is also the location where $u_{\mathrm{g, ref}}$ and the corresponding strain rate are determined. The most accurate extrapolation to $s_{\mathrm{L}}$ is obtained when using the gas velocity, which is the reason why $u_{\mathrm{g}}$ is used instead of $u_{\mathrm{p}}$. For an increased mass-flux the maximum strain rate will increase and the particle inertia effects are more severe, see \cref{fig:strain-Qmax,fig:u_ref-Qmax}. Furthermore, the flame will move towards the stagnation point and at a high enough mass-flux the flame will quench. Figure~\ref{fig:K-sL-Qmax} shows that $u_{\mathrm{g, ref}}$ varies in a parabolic way with the stretch rate. Hence, by parabolic extrapolating the values of $u_{\mathrm{g, ref}}$ determined at the various strain rates, the intercept on the ordinate ($a_{\mathrm{g}}$~=~0), should give the laminar burning velocity $s_{\mathrm{L}}$ for the one-dimensional stretch-free flame after correcting for gas expansion effects. This correction follows from conservation of mass for a flame with zero strain:
\begin{align}
	u_{\mathrm{g}}^u\rho_{\mathrm{g}}^u  + u_{\mathrm{p}}^u\rho_{\mathrm{p}}^u&=  u_{\mathrm{g}}^b\rho_{\mathrm{g}}^b  + u_{\mathrm{p}}^b\rho_{\mathrm{p}}^b,\\
	s_L &= \frac{\rho^b_g u_{\mathrm{g, ref}}^b + u_{\mathrm{p}}^b\rho_{\mathrm{p}}^b -u_{\mathrm{p}}^u\rho_{\mathrm{p}}^u}{\rho_g^u}.
\end{align}
Here use is made of the fact that for a free flame the unburned gas velocity is equal to the laminar burning velocity, $u_{\mathrm{g}}^u = s_L$, and that the velocity on the burned side is equal to the reference velocity $u_{\mathrm{g, ref}}^b = u_{\mathrm{g}}^b$. 
The maximum heat release is not something that can be measured very easily in practice, but a maximum light intensity can be measured instead. On top of that, in order to perform the gas expansion correction, the particle mass flux must be known very accurately, which is also still a challenge.

\begin{figure}
	\centering	
	\subfigure[]{\includegraphics[width=0.48\textwidth]{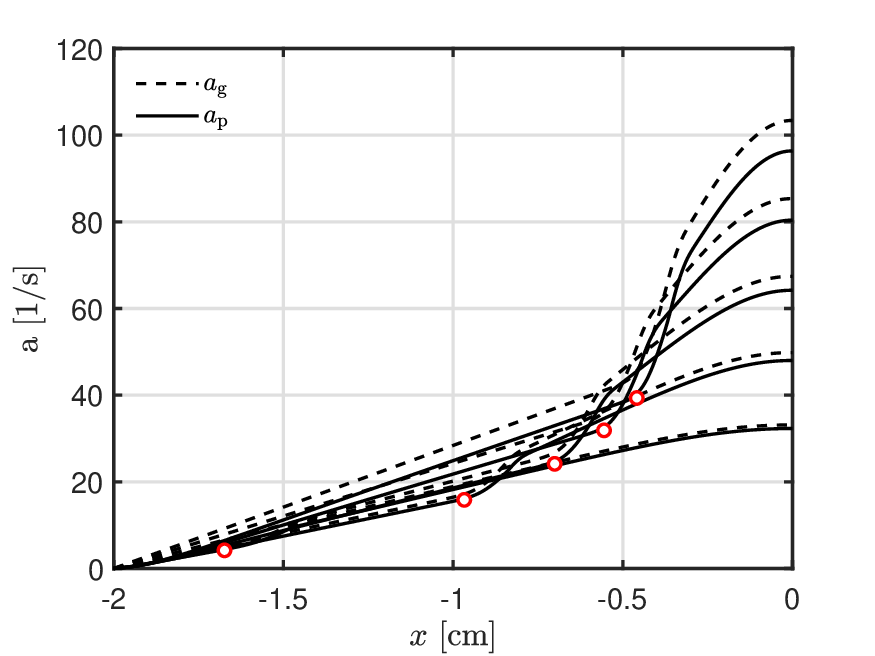}\label{fig:strain-u_min}}	
	\quad
	\subfigure[]{\includegraphics[width=0.48\textwidth]{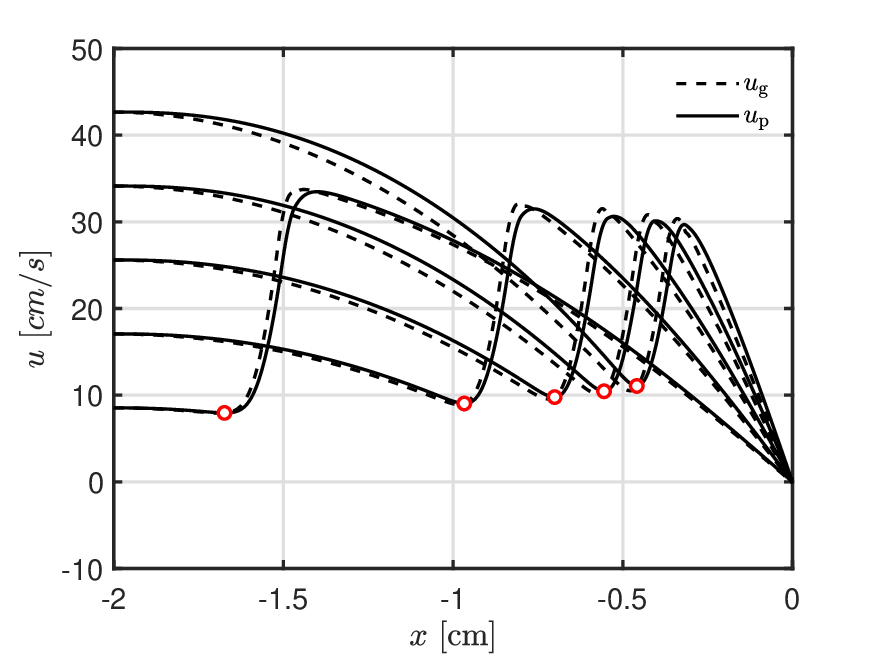}\label{fig:u_ref-u_min}}
	\quad
	\subfigure[]{\includegraphics[width=0.48\textwidth]{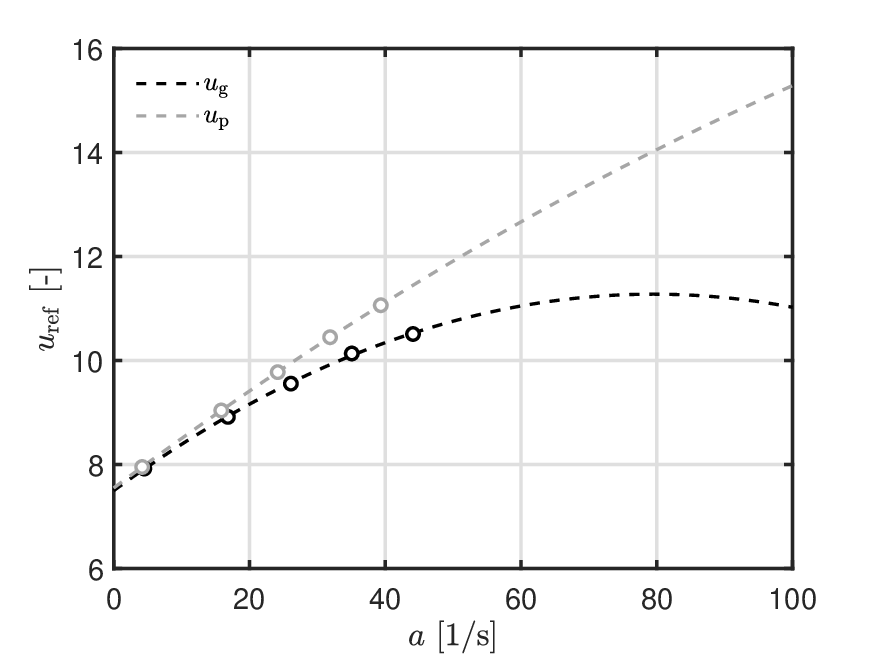}\label{fig:K-sL-u_min}}	
	\caption{$s_{\mathrm{L}}$ determination at the location of $u_{\mathrm{min}}$ method. a) Gas (dashed) and particle (solid) strain profiles. b) Gas (dashed) and particle (solid) velocity profiles. c) $u_{\mathrm{ref}}$ as function of strain rate. Simulations are performed at $\phi$~=~0.6 and $\dot{m}$~=~0.01~-~0.05~g/cm$^2$s, with an interval of 0.01~g/cm$^2$s. The red symbols indicate the location of $u_{\mathrm{min}}$.} \label{fig:sL-u_min}
\end{figure} 

\begin{figure}
	\centering
	\includegraphics[width=260pt]{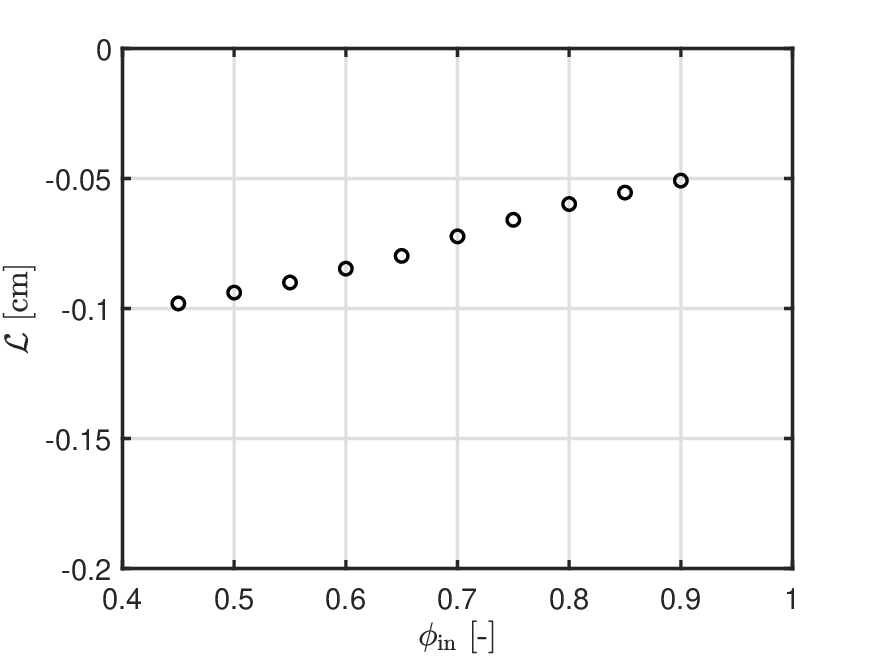}
	\caption{Markstein length $\mathcal{L}$ for different fuel equivalence ratios.}
	\label{fig:Markstein}
\end{figure}

The other option we will look into is the minimum velocity variant, where velocity and strain rate are measured at the location of minimum particle velocity, $u_{\mathrm{p, ref}}$. For this variant we will use the particle velocity instead of the gas velocity, as this is what can actually be measured in experiments. In Fig.~\ref{fig:sL-u_min} the strain rate, velocity profiles and the effect of strain on the velocity at the location of minimum velocity for $\phi_{\textrm{in}}$~=~0.6 and mass-flux $\dot{m}$~=~0.01~-~0.05~g/cm$^2$ are shown.  Figure.~\ref{fig:K-sL-u_min} shows that $u_{\mathrm{p, ref}}$ varies parabolic with the stretch rate. Similar to $Q_{\mathrm{max}}$ method, we find that by parabolic extrapolating the values of $u_{\mathrm{p, ref}}$ determined at the various strain rates, the intercept on the ordinate ($a$~=~0), gives the laminar burning velocity $s_{\mathrm{L}}$ for the one-dimensional stretch-free flame. In Fig.~\ref{fig:K-sL-u_min} the minimum gas velocity and it's corresponding strain rate are also shown, where it can be seen that the dependency of the minimum velocity on the strain rate is different from that of the particles. Nevertheless, the laminar burning velocity determined with $u_{\mathrm{g, ref}}$ is similar to the one determined with $u_{\mathrm{p, ref}}$. In Fig.~\ref{fig:Markstein} the Markstein lengths, $\mathcal{L}$ are shown for a range of fuel-equivalence ratios, which shows a roughly linear dependency. The Markstein length, is a measure of the response of the flame to stretch \cite{TSENG1993}, which relates the flame propagation rate $s^*_{\mathrm{L}}$ to the laminar (unstretched) flame speed and the strain rate according to:
\begin{align}
	s^*_{\mathrm{L}} &= s_{\mathrm{L}} - \mathcal{L}a.
\end{align}

\begin{figure}
	\centering
	\includegraphics[width=300pt]{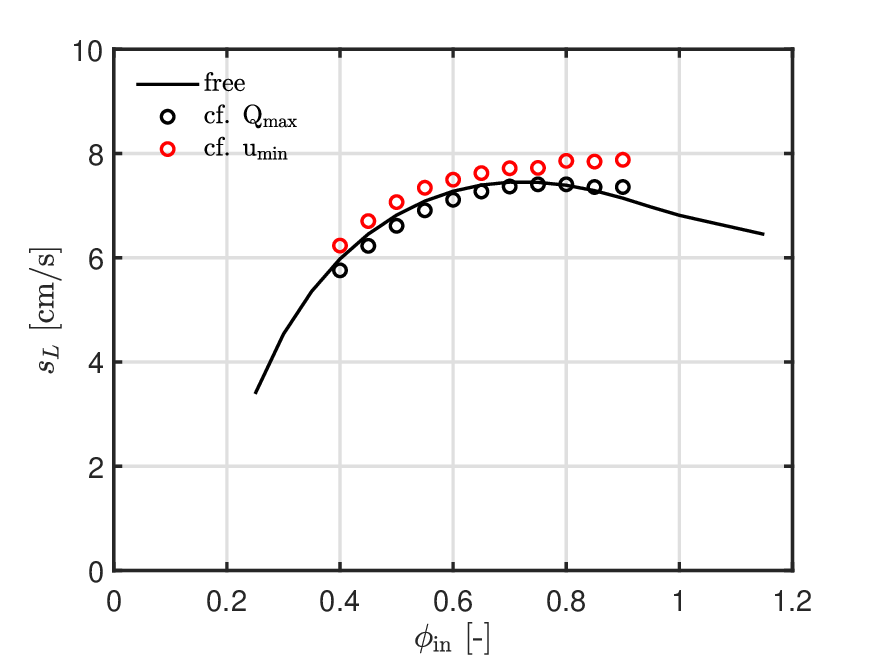}
	\vspace{10 pt}
	\caption{Laminar flame speeds of 10 \textmu m iron-dust-air mixtures (black solid) free flame, (black symbols) counter flow $Q_{\mathrm{max}}$ method, and (red symbols) counter flow $u_{\mathrm{min}}$ method.}
	\label{fig:phi_in-sL_ref}
\end{figure}

The results of $s_{\mathrm{L}}$ predictions from the free flame and stagnation flame are presented and compared in Fig.~\ref{fig:phi_in-sL_ref}. Overall, the flame speed predictions of both methods are good, but the $Q_{\mathrm{max}}$ method performs slightly better. Also, for $\phi_{\mathrm{in}} >$~0.8 a larger discrepancy between the free flame $s_{\mathrm{L}}$ prediction and the $u_{\mathrm{min}}$ method stagnation flame predictions is observed. For these fuel equivalence ratios, fewer data points were available at very low strain rates, which probably influenced the extrapolation to zero strain rate. 

Based on the two methods shown here, we would recommend to use the $u_{\mathrm{min}}$ method as this method is in good agreement with the free flame $s_{\mathrm{L}}$ prediction and more feasible from an experimental point of view. 

%% file: Conclusions.tex
\section{Conclusions}\addvspace{10pt}
In this work, a detailed 1D strain model is derived, assuming Re to be small. This detailed model is used to investigate the influence of strain on iron dust counter-flow flames. Flame structures of a free flame, a counter-flow flame where gas flow strain equals particle flow strain and a counter-flow flame where the particle flow strain is resolved are compared to each other. Thereafter, the influence of strain on various particle sizes is briefly discussed. Lastly, a study is performed on the prediction of $s_{\mathrm{L}}$ using two variants of one method: the $Q_{\mathrm{max}}$ variant and the $u_{\mathrm{min}}$ variant.

First, it is concluded that preferential diffusion, due to the lack of diffusion in the fuel, is always at play in (iron) dust flames. This implies that the specific enthalpy and elemental fractions do not remain constant throughout the flame front. This is a relevant observation, as for positive stretch the mixture becomes richer while the specific enthalpy decreases. However, this does not mean that the flame temperature reduces since the mixture $c_{\mathrm{p}}$ decreases. 

The importance of solving a particle flow strain model instead of assuming particle flow having an equal strain as the gas flow is demonstrated: If the particle flow strain is assumed equal to the gas strain, the burned side of the flame actually becomes leaner instead of richer. The 'inertia effect' of the particles, which we showed dominant over preferential diffusion, is neglected when simply assuming the two strain rates to be equal to each other. 

It is shown in the third subsection of the results that particle flow strain effects are of importance when performing experiments that are prone to strain. In these experiments each particle size experiences a different strain rate and thereby the PSD at the flame-front will be different than at the inlet. Future numerical studies could investigate this. 

Finally, in the last subsection of the results a study is performed on the prediction of $s_{\mathrm{L}}$ with two different methods, the $Q_{\mathrm{max}}$ method and the $u_{\mathrm{min}}$. Both methods are in good agreement with the free flame $s_{\mathrm{L}}$ prediction, but from an experimental point of view the $u_{\mathrm{min}}$ method is easier to use and therefore recommended. 